\documentclass[10pt]{article}
\usepackage{amssymb}
\usepackage{epsfig}
\title{Decay Process $\sigma \rightarrow \pi \pi$ and Chiral Phase Transition}
\author{Zhu Xianglei, Zhuang Pengfei\\Department of Physics, Tsinghua University, Beijing 100084, China}
\begin{document}
\maketitle
\begin{abstract}
{\it The decay process $\sigma\rightarrow\pi\pi$ at high temperature and density and its relation with chiral phase
transition are discussed in the framework of the NJL model. The decay rate for the process $\sigma\rightarrow\pi\pi$
is calculated in the whole temperature and density region. The contribution of the final state pion statistics
to the decay rate is discussed. The maximum decay rate at different chemical potential is computed. Finally, we investigate
the relation between the starting point of the decay process and the critical point of the first-order chiral phase transition.}
\end{abstract}
\section{Introduction}
It's generally believed \cite{muller} that there are two kinds of QCD phase transitions in relativistic heavy ion collisions. One of them is related to the deconfinement
process of quarks, and the other is about chiral symmetry restoration. When we investigate chiral properties, $\sigma$ meson has to be
included \cite{rho}. While the $\sigma$ mass and its width are very large in the vacuum,
around the chiral phase transition point $\sigma$ becomes light and its width is also  very small \cite{zhuang1}. Therefore there will be numerous $\sigma$'s in the early stage of a relativistic heavy ion collision if the system undergoes a chiral phase transition. When the temperature and density of the system
drop down, $\sigma$ mass increases but pion mass keeps as a constant. When the mass condition
$m_{\sigma}=2m_{\pi}$ is reached, the process $\sigma\rightarrow\pi\pi$ starts immediately. From these considerations, if chiral phase transition happens in the system, the numerous $\sigma$'s around the phase-transition
point will affect remarkably the final state pions in magnitude and in momentum distribution \cite{stephanov}.

Nambu--Jona-Lasinio(NJL) model \cite{nambu} is often used to
investigate spontaneous chiral symmetry breaking in the vacuum and
chiral symmetry restoration at finite temperature and density
\cite{vogl}. Although the model has no confinement mechanism in
the vacuum and therefore can not be used to study deconfinement
\cite{vogl}, it describes chiral properties very well
\cite{quack}. In this article we calculate in the framework of the
NJL model the decay rate for $\sigma\rightarrow\pi\pi$ and
investigate its relation with the chiral phase transition.

\section{The decay rate for $\sigma\rightarrow\pi\pi$}

The Lagrangian density of SU(2) NJL model is
\begin{equation}
L_{NJL}=\bar{\psi}(i \gamma^\mu \partial_{\mu}-m_0)\psi+G[(\bar{\psi}\psi)^2+(\bar{\psi}i \gamma_5\tau \psi)^2],
\label{eqn1}
\end{equation}
where $\psi$ and $\bar{\psi}$ are the quark fields, $\tau$ is the isospin generator, G is the coupling constant
with dimension $GeV^{-2}$, and $m_0$ is the current quark mass.

\begin{figure}[ht]
\hspace{6.3cm}
\parbox{8cm}{
     \epsfig{file=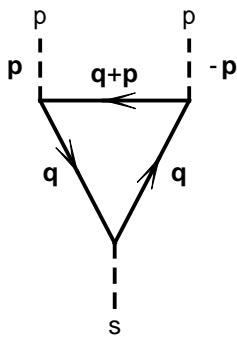}}
\caption{\it Feynman diagram for the process $\sigma\rightarrow\pi\pi$ to the lowest order. The solid lines denote quarks and anti-quarks, the dashed lines denote mesons.}
\label{fig1}
\end{figure}

Taking quark propagator in the mean field approximation and meson propagator in the RPA approximation \cite{vogl},
we get the lowest order Feynman diagram for the decay process $\sigma \rightarrow \pi \pi$ sketched in Fig.~\ref{fig1}. The decay
rate for the process at finite temperature and density is
\begin{eqnarray}
\Gamma_{\sigma \rightarrow 2\pi}(T,\mu)&=&\Gamma_{\sigma \rightarrow 2\pi^0}(T,\mu)+\Gamma_{\sigma \rightarrow \pi^+\pi^-}(T,\mu)\nonumber\\
&=&\frac{3}{8\pi}\frac{\sqrt{m_\sigma^2/4-m_\pi^2}}{m_\sigma^2}g_\sigma^2 g_\pi^4|A_{\sigma \pi \pi}(T,\mu)|^2[1+2f_B(\frac{m_\sigma}{2})],
\label{eqn2}
\end{eqnarray}
where the triangle factor $A_{\sigma\pi\pi}$ is defined by \cite{hufner}
\begin{eqnarray}
A_{\sigma \pi \pi}(T,\mu)&=&4m_q N_c N_f \int \frac{d^3\vec{q}}{(2\pi)^3}\frac{f_F(E_q-\mu)-f_F(-Eq-\mu)}{2E_q}\times \nonumber\\
&&\frac{8(\vec{q}\cdot \vec{p})^2-(2m_\sigma^2+4m_\pi^2)\vec{q}\cdot \vec{p}+m_\sigma^4/2-2m_\sigma^2E_q^2}{(m_\sigma^2-4E_q^2)((m_\pi^2-2\vec{q}\cdot \vec{p})^2-m_\sigma^2E_q^2)}.
\label{eqn3}
\end{eqnarray}
In Eq.~(\ref{eqn2}), T and $\mu$ are temperature and bayonic chemical potential respectively, $m_\sigma$ and $m_\pi$ are $\sigma$ mass and $\pi$ mass,
$g_\sigma$ and $g_\pi$ are the coupling constants of $q\bar{q}\sigma$ and $q\bar{q}\pi$ interactions. The temperature and density dependence of these quantities can be found in Ref.~\cite{zhuang1}.
$f_B(x)=(e^{x/T}-1)^{-1}$ in Eq.~(\ref{eqn2}) is the Bose-Einstein distribution function for the final state pions. In Eq.~(\ref{eqn3}), $N_c=3$ and $N_f=2$ are the quark color and flavor
degrees of freedom, $f_F(x)=(e^{x/T}+1)^{-1}$ is the Fermi-Dirac distribution function for quarks, $E_q=\sqrt{m_q^2+q^2}$ is the quark energy. The temperature and density dependence of
quark mass $m_q$ can also be found in Ref.~\cite{zhuang1}.

\section{The relation with the chiral phase transition}

We first discuss the medium effect for the final state pions in the decay process. The temperature dependence($\mu=0$) of the process $\sigma \rightarrow \pi \pi$ was considered in Ref.~\cite{zhuang2}. The difference between Ref.~\cite{zhuang2} and Eq.~(\ref{eqn2}) is in the consideration
of the statistics for the final state pions. It can be proved \cite{hatsuda} that the temperature and density effect of two final state mesons can be represented as a  statistical factor
$(1+f_1+f_2)$. In the case of $\sigma \rightarrow \pi \pi$, the correct statistical factor is therefore $(1+2f_B(\frac{m_\sigma}{2}))$ in Eq.~(\ref{eqn2}), but
not $(1+f_B(\frac{m_\sigma}{2}))^2$ in Ref.~\cite{zhuang2}. This difference in the final state statistics will result in an essential change in the behavior of the decay rate at the critical point of chiral phase transition.

\begin{figure}[t!]
\hspace{3.5cm}
\parbox{11cm}{
     \epsfig{file=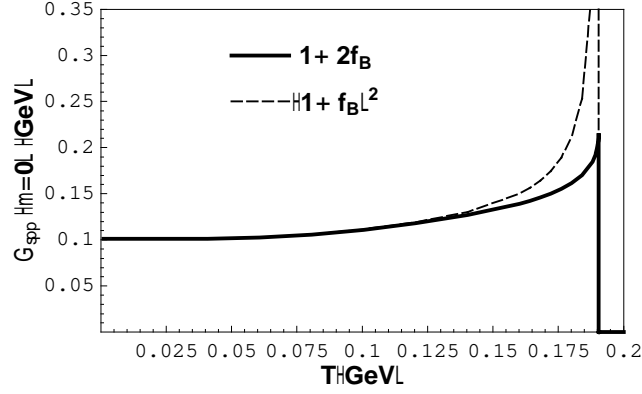}}
\caption{\it The decay rate with different statistical factors in the chiral limit($\mu=0$).}
\label{fig2}
\end{figure}

The parameters of the NJL model in chiral limit can be determined by the quark condensate density $<\bar{u}u>=<\bar{d}d>=(-0.25GeV)^3$ and the pion decay constant $f_{\pi}=0.093GeV$ in
the vacuum$(T=0, \mu=0)$, they are the coupling constant $G=5.02GeV^{-2}$ and the quark momentum cutoff $\Lambda=0.653GeV$.
The decay rate $\Gamma_{\sigma \pi \pi}$ as a function of temperature T $(\mu=0)$ is displayed in Fig.~\ref{fig2}.
The decay begins and reaches suddenly the maximum decay rate at the critical  temperature $T_c=190MeV$.
The solid line represents the decay rate with statistical factor $(1+2f_B)$ and the dashed line represents
that with $(1+f_B)^2$. It can be found in Fig.~\ref{fig2} that, there is almost no difference between the two factors at low
temperature. However, when the temperature approaches to the critical point of the  second-order chiral phase transition, the first
statistical factor makes the decay rate finite at the critical point, but the later one makes it infinite.
From this we can conclude that the chiral phase transition leads to a strong pion  enhancement, but the singularity in the decay rate comes from the wrong consideration of the medium effect for the final state pions. By further computation, we can get the analytic expression of the decay rate at the critical point,
\begin{equation}
\Gamma_{\sigma \pi \pi}(T \rightarrow T_c^-)=\frac{3\pi T_c}{N_c N_f}\frac{1}{\int_0^{\Lambda}dq\frac{1}{q}[f_F(-E_q-\mu)-f_F(E_q-\mu)]}.
\label{eqn4}
\end{equation}

\begin{figure}[t!]
\hspace{2.5cm}
\parbox{11.1cm}{
     \epsfig{file=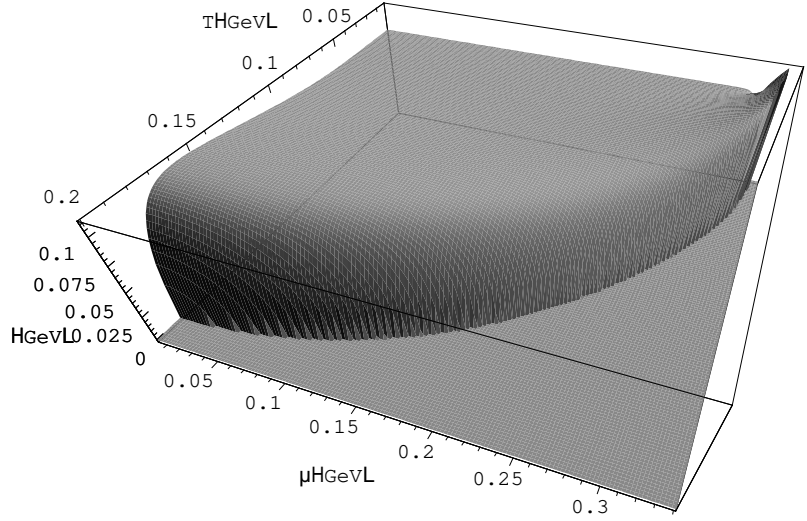}}
\caption{\it The decay rate in the whole T-$\mu$ region.}
\label{fig3}
\end{figure}

The pion mass is not zero in the real physical world. The explicit chiral symmetry breaking, namely $m_0\ne 0$, must be considered.
In this case, the three parameters in the NJL model can be determined by the quark condensate, the pion coupling constant and the pion mass $m_\pi=0.134GeV$
in the vacuum, they are $G=4.93GeV^{-2}$, $\Lambda=0.6535GeV$ and $m_0=0.005GeV$. Fig.~\ref{fig3} shows the decay rate in the whole
temperature and density region beyond chiral limit. In the vacuum, the decay rate is about 0.09GeV.
Keeping T=0 and increasing $\mu$, the decay rate behaves as a constant until $\mu$ is close to the critical point of the first-order
chiral phase transition. The decay rate decreases slightly at $\mu=0.3 GeV$, then increases and reaches the maximum value at
the critical point. Finally it jumps down to zero at the critical point. In the high density region where the first-order phase transition
exists, this jump retains at the chiral phase transition points. Whether the decay rate jumps to zero
will be discussed in a while. When the chemical potential $\mu$ is not so high that first-order phase transition can't happen($\mu<0.327GeV$),
with increasing temperature the decay rate rises gradually from its vacuum value to the maximum at some temperature, and this temperature decreases with increasing  chemical potential. After reaching the maximum, the decay rate drops down rapidly
but continuously to zero at the threshold temperature defined by  $m_{\sigma}(T,\mu)=2m_{\pi}(T,\mu)$. The fact that the maximum value
of $\Gamma$ is not in the vacuum but near the threshold point is
attributed to the contribution
of the final state pion statistics which is important only at high temperature.

\begin{figure}[t!]
\hspace{4.4cm}
\parbox{11cm}{
     \epsfig{file=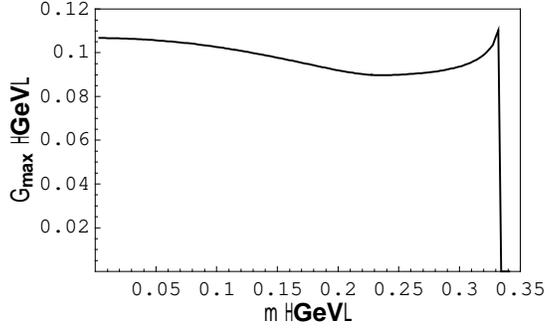}}
\caption{\it The maximum decay rate at different chemical potential $\mu$.}
\label{fig4}
\end{figure}

\begin{figure}[t!]
\hspace{3.5cm}
\parbox{10cm}{
     \epsfig{file=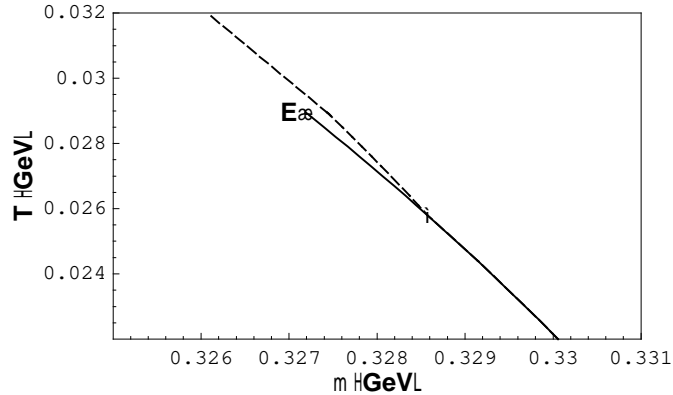}}
\caption{\it The solid line denotes the first-order chiral phase transition line around endpoint E, the dashed line denotes the threshold point for the decay at different $\mu$.}
\label{fig5}
\end{figure}

For each chemical potential $\mu$, we can extract a maximum decay rate $\Gamma_{max}$ from $\Gamma(T,\mu)$. Fig.~\ref{fig4}
shows the $\Gamma_{max}$ as a function of $\mu$. When chemical potential is very low or very high, namely in the
limit of high temperature or high density, $\Gamma_{max}$ is larger than the values in the middle region. However, the variation is not big, we can consider $\Gamma_{max}$ as a constant approximately.

In the case of explicit chiral symmetry breaking, the second-order chiral phase transition at high temperature
in the chiral limit becomes a smooth crossover, while the first-order phase transition at high density still exists with a new critical point \cite{zhuang3}. In Fig.~\ref{fig5}, the solid line represents the critical temperature $T_c$
of the first-order chiral phase transition at different chemical potential $\mu$, the point E is the endpoint of
the phase transition. As discussed above, the threshold point of the decay $\sigma \rightarrow \pi
\pi$ is defined by the mass equation $m_{\sigma}(T_d,\mu_d)=2m_\pi(T_d,\mu_d)$. The
question is if the threshold point coincides with the critical point, or if $T_d$ equals to $T_c$.

The dashed line in Fig.~\ref{fig5} represents the threshold temperature $T_d$ at different $\mu$. It is easy to see that, $T_d$ is larger than $T_c$ in the density region
near the endpoint E. The difference between $T_d$ and $T_c$ is maximum at E, and decreases with increasing chemical potential. At some chemical potential it vanishes. Therefore, in the first-order phase transition region near the endpoint E, the decay rate jumps at the phase transition point,
but it does not jump to zero but to a finite value. After this jump the decay rate  drops continuously to zero at $T_d$. In the rest first-order phase transition region where the critical point and the threshold point coincide,
the decay rate jumps to zero. Although the decay $\sigma \rightarrow \pi \pi$ happens
before the phase transition around the endpoint E, our numerical calculations show that only about $1\%$ of $\sigma$'s decay before the phase transition.

\section{Conclusions}

Whether the chiral phase transition happens in relativistic heavy ion collisions can be manifested by the number and momentum spectra of final state pions. In mean-field approximation, the chiral symmetry restoration changes mainly the masses of particles , and in turn the thresholds of particle reactions. Since the different temperature and density dependence of $\sigma$ and $\pi$,
the process $\sigma \rightarrow \pi \pi$ at finite temperature and density changes the properties of final state pions remarkably. Around the chiral phase transition point $\sigma$ becomes very light and is easy to be generated. These produced $\sigma$'s will decay
into two pions when the decay condition $m_\sigma = 2 m_\pi$ is satisfied. When we consider the medium effect for the final state pions correctly, the enhancement will not become unphysically infinite. The almost chemical potential independent maximum decay rate means that the chiral properties reflected by
$\sigma \rightarrow \pi \pi$ can be measured in relativistic heavy ion collisions at
different energy. While in the chiral limit the threshold point of
$\sigma \rightarrow \pi \pi$ coincides with the critical point of the chiral phase transition, in the case of explicit chiral symmetry breaking,
the decay may happen before the chiral phase transition. Our numerical results showed in this paper depend certainly on the NJL model, but we believe that the above qualitative conclusions are model independent.

\section*{Acknowledgements}

The authors are grateful to Huang Mei and Yang Zhenwei for their valuable discussions. This work was supported
in part by the NSFC (19925519), the 973 project (G2000077407) and the National Laboratory of Heavy Ion Physics in Lanzhou.

\end{document}